\documentclass[a4paper,11pt]{article}

\usepackage{jcappub} 
\usepackage[export]{adjustbox} 
\usepackage[utf8]{inputenc}
\usepackage{color}
\usepackage{graphicx}
\usepackage{amsmath,amsfonts,amssymb}
\usepackage{acronym}
\usepackage{xspace}
\usepackage{multirow}
\usepackage{tabularx}
\usepackage{orcidlink}
\usepackage{ragged2e}
\usepackage{enumitem}
\usepackage{setspace}
\usepackage{caption}
\usepackage{subcaption}
\usepackage{booktabs}
\usepackage{comment}
\usepackage{braket}
\DeclareMathOperator{\Tr}{Tr}

\usepackage{listings}
\usepackage[ruled,vlined]{algorithm2e}

\SetCommentSty{mycommfont}

\newcommand{\bea}{\begin{equation}\begin{aligned}} 
\newcommand{\eea}{\end{aligned}\end{equation}}
\newcommand{\be}{\begin{equation}}
\newcommand{\ee}{\end{equation}}

\definecolor{rossocorsa}{rgb}{0.83, 0.0, 0.0}
\definecolor{tardisblue}{rgb}{0.0, 0.18, 0.53}
\definecolor{TardisBlueAcceso}{RGB}{0, 80, 200}

\title{Quantum production of gravitational waves after inflation}

\author[a,b]{Alina Mierna,}
\author[a,b,c]{Gabriele Perna,}
\author[a,b,d,f]{Sabino Matarrese,}
\author[a,b,d,f]{Nicola Bartolo,}
\author[g,h]{Angelo Ricciardone}

\affiliation[a]{Dipartimento di Fisica e Astronomia ``G. Galilei'',
Universit\`a degli Studi di Padova, via Marzolo 8, I-35131 Padova, Italy}
\affiliation[b]{INFN, Sezione di Padova,
via Marzolo 8, I-35131 Padova, Italy}
\affiliation[c]{Institute for Theoretical Physics, Leibniz University Hannover, Appelstraße 2, 30167 Hannover, Germany}
\affiliation[d]{INAF- Osservatorio Astronomico di Padova, \\ Vicolo dell’Osservatorio 5, I-35122 Padova, Italy}
\affiliation[f]{Gran Sasso Science Institute, Viale F. Crispi 7, I-67100 L’Aquila, Italy}
\affiliation[g]{Dipartimento di Fisica “Enrico Fermi”, Universit\`a di Pisa, Pisa I-56127, Italy}
\affiliation[h]{INFN sezione di Pisa, Pisa I-56127, Italy}

\abstract{A variety of mechanisms in the early Universe lead to the generation of gravitational waves (GWs). We introduce here a novel source of GWs generated by vacuum fluctuations after inflation. Given that gravitons are minimally coupled particles, their quantum creation takes place during inflation, but is absent in an unperturbed Universe during the radiation-dominated epoch, since they behave as conformally coupled particles. However, the presence of inhomogeneities breaks the conformal flatness of the metric, allowing scalar metric perturbations to induce the quantum production of gravitons. We compute the resulting GW spectrum from this mechanism for different models of the primordial scalar power spectrum. We find that this GW signal peaks around the GHz frequency range, distinguishing it from other astrophysical and cosmological backgrounds and underscoring the need for detectors sensitive to these high frequencies.}

\begin{document}
\maketitle
\flushbottom
\section{Introduction} 

In an expanding universe, particle creation can take place solely due to coupling to gravity~\cite{Parker:1968mv, Parker:1969au, Parker:1971pt, Ford:2021syk, Kolb:2023ydq, Capanelli:2024pzd}. The cosmic expansion, in fact, induces a mixing of the positive and negative frequency modes, leading to the quantum creation of particles, and this mixing is regulated by a Bogoliubov transformation, which connects the set of ladder operators in the asymptotic past and future. Minkowski spacetime is static and therefore no particle production occurs: a positive frequency mode in the past will remain the positive frequency mode in the future. On the other hand, in curved spacetime there is no general definition of the vacuum state. Since the spatially flat Friedmann-Lema\^itre-Robertson-Walker (FLRW) spacetime is related to Minkowski spacetime by a conformal transformation, no massless conformally coupled particles can be created in this way. However, the conformal invariance is broken by the presence of inhomogeneities, allowing for the production of massless conformally coupled particles, such as photons~\cite{Zeldovich:1977vgo,Birrell:1979pi, Cespedes:1989kh,Maroto:2000zu}. The cosmological gravitational particle production (CGPP) has recently attracted attention as an appealing mechanism for dark matter generation, since it does not require the interaction of dark matter with fields other than gravity~\cite{Maleknejad:2024ybn, Maleknejad:2024hoz, Garani:2024isu, Garani:2025qnm, Belfiglio:2025chv}. Gravitons are minimally coupled and created even in the spatially flat FLRW spacetime, as it happens during inflation. Conversely, during the radiation-dominated epoch, the Ricci scalar vanishes and gravitons behave as conformally coupled particles. Therefore, the presence of inhomogeneities is necessary to generate gravitational waves (GWs) during the radiation-dominated epoch. 

GWs provide a unique probe of the early Universe, since they decouple at the Planck scale and travel freely afterwards. The PTA collaboration has recently reported evidence for the detection of a Gravitational Wave Background (GWB) in the nHz frequency range~\cite{NANOGrav:2023gor, EPTA:2023fyk, Reardon:2023gzh,Xu:2023wog}. 
A variety of possible sources of the detected signal has been analyzed of both astrophysical and cosmological origin~\cite{NANOGrav:2023hvm, NANOGrav:2023hfp, EPTA:2023xxk, Franciolini:2023pbf, Franciolini:2023wjm, Ellis:2023tsl, Vagnozzi:2023lwo, Figueroa:2023zhu,Domenech:2024rks}.
In this work, we propose a novel mechanism that can generate GWs from vacuum fluctuations after inflation. We consider the CGPP of GWs, induced by the scalar perturbations of the metric during the radiation-dominated epoch. 
The GWs produced by the coupling of scalar and tensor perturbation at second order in the radiation-dominated epoch have been studied in~\cite{Bari:2023rcw, Picard:2023sbz}. Such a process is classical and requires a pre-existing GW signal to induce the scalar-tensor mixing. Here instead we consider the quantum production of gravitons by the perturbed background metric, which is independent of the primordial GWs. 
As a matter of fact, any primordial GWs will lead to the stimulated emission process~\cite{Agullo:2011xv,Ota:2025yeu}, since gravitons already present in the initial state will amplify the quantum generated signal. 

It is important to note that this process is semiclassical, in the sense that gravitons are quantized, but the perturbed background metric is treated as classical. We consider only the scalar background perturbations neglecting the tensor modes as they are expected to be subdominant in standard models of inflation.

In the following, we adopt natural units $c = 1 $ and $  H_0 = 67.66\, {\rm km/s/Mpc} \simeq 2.25 \cdot10^{-4} {\rm Mpc}^{-1}$.

\section{GWs from vacuum fluctuations}
The perturbed FLRW metric in the Poisson gauge reads
\begin{equation}
\begin{split}
    ds^2 &= a^2(\eta)\left[-e^{2\Psi}d\eta^2 +e^{-2\Phi}\left(\delta_{ij} + h_{ij}\right)dx^idx^j\right]\,,  
\end{split}
\end{equation}
where we consider the tensor perturbations up to second order $h_{ij} = h_{ij}^{(1)} + \frac{1}{2}h_{ij}^{(2)}$. 
The Einstein equations for GWs at second order read
\begin{equation}
    \mathcal{T}^{lm}_{ij} G^{(2)}_{lm} = 8 \pi G  \mathcal{T}^{lm}_{ij} T^{(2)}_{lm}\,,
\end{equation} 
where $\mathcal{T}^{lm}_{ij}$ is the transverse-traceless projector, see e.g. \cite{Bari:2023rcw, Picard:2023sbz}.
In Fourier space, the tensor and scalar perturbations can be written as 
\begin{equation}
\begin{split}
    h_{ij}(\Vec{x},\eta) &=  \sum_{\lambda}\int \frac{d^3k}{(2\pi)^3}h_{\lambda, \Vec{k}}(\eta) \epsilon^\lambda_{ij}(\hat{k})e^{i\Vec{k}\cdot\Vec{x}}\,,\\
    \Psi(\Vec{x},\eta) &= \int \frac{d^3q}{(2\pi)^3} \Psi(\Vec{q},\eta) e^{i\Vec{q}\cdot\Vec{x}}  \,,    
\end{split}
\end{equation}
where $\epsilon^\lambda_{ij}(\hat{k})$ is the polarization tensor and 
$\lambda$ represent the two GW polarizations.
The equation of motion for each GW polarization is equivalent to the equation of motion for a minimally coupled scalar field \cite{Ford:1977dj}. In the following, we will work in terms of $\gamma_{\lambda, \Vec{k}}$, defined as \cite{Caprini:2018mtu} 
\begin{equation}
    h_{\lambda, \Vec{k}} =    \frac{\sqrt{2} }{m_{\rm Pl} a(\eta)}\gamma_{\lambda, \Vec{k}} \,,
\end{equation}
where $m_{\rm Pl}=1/\sqrt{8 \pi G}$ is a reduced Planck mass.
Neglecting the anisotropic stress, $\Phi=\Psi$, we can write the equation of motion for GWs sourced by scalar perturbations as
\begin{equation}
\label{Eq:inhomo}
    \begin{split}        \gamma_{\lambda,\Vec{k}}^{\prime\prime} + \left(k^2 - \frac {a^{\prime\prime}}{a} \right)\gamma_{\lambda,\Vec{k}} =  J_{\lambda}(\Vec{k},\eta)\,,
    \end{split}
\end{equation}
where primes denote differentiation w.r.t. the conformal time $\eta$, 
and the source term $J_{\lambda}(\Vec{k},\eta)$ is a quadratic combination of the first-order scalar and tensor perturbations, 
\begin{equation}
    \begin{split} J_{\lambda}(\vec{k},\eta)= &\sum_{\sigma}\int\frac{d^3q}{(2\pi)^3}\epsilon_{ij}^{\sigma}(\hat{q})\epsilon^{ij,*}_\lambda(\hat{k})\gamma_{\sigma,\vec{q}} \left[\left(-4\vec{q}\cdot\vec{k}-2(1-w)(\vec{k}-\vec{q})^2\right) \Psi(|\vec{k}-\vec{q}|,\eta)\right.\\
    & \left. + 2\mathcal{H}(1+3w)\Psi^\prime(|\vec{k}-\vec{q}|,\eta)\right]=\sum_{\sigma}\int\frac{d^3q}{(2\pi)^3} Q_{\lambda,\sigma}(\vec{q},\vec{k}) I(\vec{q},\vec{k},\eta) \gamma_{\sigma,\vec{q}}
    \end{split}
\end{equation}
with $ Q_{\lambda,\sigma}(\vec{q},\vec{k}) = \epsilon_{ij}^{\sigma}(\hat{q})\epsilon^{ij,*}_\lambda(\hat{k})$ and $w$ the equation of state parameter. We consider that the spacetime is asymptotically flat (Minkowski-like) both at early and late times, i.e., in the  limit $\eta \rightarrow -\infty$ and  $\eta\rightarrow + \infty$ respectively. Promoting the field to a quantum operator in the asymptotic past and future, the solutions read
\begin{equation}
\begin{split}
      & \gamma_{\lambda,\vec
    {k}}^{\rm in}(\eta\rightarrow-\infty) = \frac{e^{-ik\eta}}{\sqrt{2k}} a_{\vec{k},\lambda} + \frac{e^{ik\eta}}{\sqrt{2k}} a_{-\vec{k},\lambda}^\dagger \, ,\\ 
    & \gamma_{\lambda,\vec{k}}^{\rm out}(\eta\rightarrow\infty) = \frac{e^{-ik\eta}}{\sqrt{2k}} \bar{a}_{\vec{k},\lambda} + \frac{e^{ik\eta}}{\sqrt{2k}} \bar{a}_{-\vec{k},\lambda}^\dagger\, ,
\end{split}
\end{equation}
where ${a}_{-\vec{k},\lambda}^\dagger$ and $a_{\vec{k},\lambda}$ are the creation and annihilation operators (the bar indicates operators associated to the asymptotic future) satisfying the commutation relations
\begin{equation}
\begin{split}
&[{a}_{\vec{k},\lambda},{a}_{\vec{q},\lambda'}] =[{a}_{\vec{k},\lambda}^\dagger,{a}_{\vec{q},\lambda'}^\dagger] = 0 \,,\\
   &[{a}_{\vec{k},\lambda},{a}_{\vec{q},\lambda'}^\dagger] = (2\pi)^3\delta_{\lambda\lambda'}\delta({\vec{k}-\vec{q}})  \, .
   \end{split}
\end{equation}
The equation can be solved in a perturbative way by the Green’s function method. In the radiation-dominated epoch, when $w=1/3$ and $a \propto \eta$, the homogeneous solutions of eq. \eqref{Eq:inhomo} behave as $e^{\pm i k \eta}$ and the corresponding Green's function is
\begin{align}
    G_{\vec{k}}(\eta,\bar{\eta}) = \frac{\sin\left[k(\eta-\bar{\eta})\right]} {k}\,.
\end{align}
The solution of eq. \eqref{Eq:inhomo} is then
\begin{equation}
    \begin{split}
 \label{solution of eom}
    \gamma_{\lambda}(\vec{k},\eta) =&\, \gamma_{\lambda}^{\rm in} (\vec{k},\eta) + \int_{\eta_{\rm in}}^{\eta} d\bar{\eta} \, G_{\vec{k}}(\eta-\bar{\eta})J_{\lambda}(\vec{k},\bar{\eta}) \\
    =&  \gamma_{\lambda}^{\rm in} (\vec{k},\eta)
    + e^{ik\eta} \int_{\eta_{\rm in}}^{\eta} d\bar{\eta} \frac{e^{-ik\bar{\eta}}}{2ik}J_{\lambda}(\vec{k},\bar{\eta})- e^{-ik\eta}\int_{\eta_{\rm in}}^{\eta} d\bar{\eta} \frac{e^{ik\bar{\eta}}}{2ik}J_{\lambda}(\vec{k},\bar{\eta}),
    \end{split}
\end{equation}
where we assume to the lowest order that $\gamma_{\lambda,\Vec{q}}$ is replaced by the initial solution $\gamma_{\lambda,\Vec{q}}^{\rm in}$ in the integrand. 
Since, after Hubble radius reentry the gravitational potentials start to decay and eventually become constant during the matter-dominated epoch, we assume that the dominant contribution to the time integral comes from the radiation-dominated epoch and therefore we set $\eta \rightarrow\infty$. For simplicity, we assume that the amount of primordial GWs is negligible\footnote{The primordial tensor power spectrum is highly suppressed in ekpyrotic and cyclic models of inflation~\cite{Baumann:2007zm}, as well as in some curvaton models~\cite{Bartolo:2007vp}.}. Therefore, at the end of inflation the gravitons can be considered to be in the Bunch–Davies vacuum state and we start the integration at $\eta_{\rm in} = 0$. Clearly, the initial state at the end of inflation is not the vacuum state and quanta are already present. We will take into account the stimulated emission of gravitons due to the presence of quanta in the initial state at the end of this letter.

The quantity of interest for GWB observations is the spectral energy density per logarithmic frequency interval that is defined as 
\begin{equation}
    \Omega_{\rm GW}(\vec{x},\eta) 
    = \frac{1}{12a^2(\eta)\mathcal{H}^2(\eta)} \langle h_{ij}^\prime(\vec{x},\eta) h^{ij\, \prime}(\vec{x},\eta) \rangle\,.
\end{equation}
where ${\cal H} = a^\prime / a$.
In Fourier space we can write
\begin{equation}
\begin{split}
    \langle h_{ij}'(\vec{x},\eta)h^{ij'} (\vec{x},\eta)\rangle & = \sum_{\lambda\lambda'} \int \frac{d^3kd^3k'}{(2\pi)^6} 
    \bra{\rm in}h_{\lambda,\vec{k}}^\prime h_{\lambda',\vec{k}'}^\prime \ket{\rm in} \epsilon_{ij}^\lambda(\hat{k})\epsilon_{ij}^{\lambda'}(\hat{k}') e^{-i(\vec{k}+\vec{k}')\cdot\vec{x}}\\
     \simeq  \sum_{\lambda\lambda'} &\frac{2}{m^2_{\rm Pl} a^2(\eta)} \int \frac{d^3kd^3k'}{(2\pi)^6}kk'
    \bra{\rm in}\gamma_{\lambda,\vec{k}} \gamma_{\lambda',\vec{k}'} \ket{\rm in} \epsilon_{ij}^\lambda(\hat{k})\epsilon_{ij}^{\lambda'}(\hat{k}') e^{-i(\vec{k}+\vec{k}')\cdot\vec{x}}\,,\\
    \end{split}
\end{equation}
where in the last line we have neglected contributions $\propto \mathcal{H}$ inside the horizon. Thus, we need now to evaluate the ``out'' number operator in the ``in'' vacuum state
\begin{equation}
   \begin{split}
           \bra{\rm in} \gamma_{\lambda,\vec{k}}\gamma_{\lambda,\vec{k}'}\ket{\rm in}  
    = & \frac{1}{2\sqrt{kk'}}\bra{\rm in} \bar{a}^\dagger_{-\vec{k},\lambda}\bar{a}_{\vec{k}',\lambda} \ket{\rm in} = \frac{1}{2\sqrt{kk'}}\sum_{\sigma}\int \frac{d^3 q}{(2\pi)^3} \beta_{-\vec{k},\vec{q},\sigma, \lambda}\beta^*_{\vec{k}',\vec{q},\sigma, \lambda}\, ,
   \end{split}
\end{equation}
where $\beta_{\vec{k},\vec{q},\sigma, \lambda}$ is the Bogoliubov coefficient associated to the negative frequency mode of $\gamma_{\lambda, \vec{k}}$ in the asymptotic future. From eq. \eqref{solution of eom}, we obtain
\begin{align}
\beta_{\vec{k},\vec{q},\sigma,\lambda}^* = & i\int_{\eta_{\rm in}}^{\eta} d\bar{\eta} Q_{\lambda,\sigma}(\vec{q},\vec{k}) I(\vec{q},\vec{k},\bar{\eta}) \left[ \frac{e^{i(k+q)\bar{\eta}}}{2\sqrt{qk}}\right]\,.
\end{align}
It is immediate to connect the Bogoliubov coefficients to the power spectrum of GWs\footnote{We specify that particle production can be derived also from the S-matrix point of view, as the decay of the perturbation to 2 particles~\cite{Cespedes:1989kh,Campos:1991ff,Parker:2009uva}.}. To compute the integrals, it is useful to perform a change of variables to $u = |\vec{k}-\vec{q}|/k$ and $v = q/k$, similarly to what is usually done for the scalar-induced
GWs~\cite{Espinosa:2018eve,Pi:2020otn,Perna:2024ehx,Kugarajh:2025rbt}. Moreover, to simplify the numerical integration, we 
perform another change of variables to $t = \left(u+v-1\right)/2$ and $s=\left(u-v\right)/2$.
\section{Results}
The power spectrum of GWs from quantum fluctuations after inflation then reads 
\begin{align}
\label{Eq:GW_Spectrum}
   &\Omega_{\rm GW}(k,\eta)  
    =  \frac{2}{3\left(2\pi\right)^2m_{\rm Pl}^2 a^4 \mathcal{H}^2} k^4\int_0^{\infty} dt \int_{-1}^{1} ds \, \sum_{\lambda,\sigma}\left|Q_{\lambda,\sigma}(u,v) \right|^2 K^2(u,v) \frac{1}{u^2} \Delta_{\zeta}(uk)\, ,
\end{align}
where $\Delta_{\zeta} (uk)$ is the primordial curvature power spectrum. The expression for the polarization tensors reads
\begin{align}
    \sum_{\lambda\sigma}|Q_{\lambda,\sigma}(\vec{q},\vec{k})|^2 & = \sum_{\lambda\sigma} \epsilon_{ij}^{\sigma}(\hat{q})\epsilon^{ij,*}_\lambda(\hat{k}) \epsilon_{lm}^{*\sigma}(\hat{q})\epsilon^{lm}_\lambda(\hat{k})\nonumber\\
    & = \left[1+6\left(\frac{1+v^2-u^2}{2v}\right)^2+\left(\frac{1+v^2-u^2}{2v}\right)^4\right]\,,
\end{align}
while for the kernel
\begin{equation}
\begin{aligned}
K^2(u,v) = \Bigg\{ 
& \left(2v \left(\frac{1+v^2-u^2}{2v}\right) + (1-w)u^2\right)\frac{\sqrt{3}}{u} \\
& \times \left[
\frac{\left(u^2 - 3 (1+v)^2\right) \coth^{-1}\left(\frac{\sqrt{3} (1+v)}{u}\right)
+ \sqrt{3} u (1+v)}{2 u^2}
\right] \\
& + 2 \Bigg[
\frac{1}{24 u^3} \Big(
u (v+1) \left(5 u^2 - 9 (v+1)^2\right) \\
& \qquad + \sqrt{3}\left(u^2 - 3 (v+1)^2\right)^2 
\coth^{-1}\left(\frac{\sqrt{3} (v+1)}{u}\right)
\Big)
\Bigg]
\Bigg\}^2\,.
\end{aligned}
\end{equation}

We consider only the GWs generated by modes that reenter the horizon during the radiation-dominated epoch. For this reason, we introduce a minimum and maximum cutoff over the internal momentum $p= |\vec{k}-\vec{q}|$, which restricts the momenta spanned by the scalar power spectrum. The maximum cutoff bounds the maximum frequency of the generated GWs. We consider as minimum value the comoving wavenumber at matter-radiation
equality, $p_{\rm min} \simeq 1.3 \cdot 10^{-2} {\rm Mpc^{-1}}$ and as a maximum value the wavenumber of the mode that exited the horizon at end of inflation,
\begin{equation}
    p_{\rm max} = a_{\rm end}H_{\rm end} = H_0 e^N_{\rm} \frac{H_{\rm end}}{H_{\rm in}}
\end{equation}
where $H_{\rm in}$ and  $H_{\rm end}$ are the Hubble parameter at the beginning and the end of inflation and inflation, $N_{\rm obs}$ 
\begin{equation}
    e^N \frac{H_{\rm end}}{H_{\rm in}} =  \frac{H_{\rm end}}{H_0}  \frac{a_{\rm end}}{a_0}
\end{equation}
Number of observable e-folds varies significantly depending on the model of inflation. On the other hand, the comoving wavenumber at the end of inflation is fixed and depends only on the details of reheating 
\begin{equation}
    p_{\rm max} = H_{\rm end}  \frac{a_{\rm end}}{a_0} \approx 3.5 \cdot 10^{23} \text{Mpc}^{-1}
\end{equation}
where we have taken the upper bound on the tensor-to-scalar ratio from \cite{Galloni:2022mok}. We also impose a condition that the comoving momentum of GWs should be smaller than the momentum of the underlying scalar perturbations, because the long-wavelength scalar perturbations are considered as a quasi  homogeneous and isotropic background from the viewpoint of produced gravitons \cite{Maleknejad:2024hoz, Maleknejad:2024ybn}.
For this reason, we do not consider GWs of momentum higher that $p_{\rm max}$. From Eq. \eqref{Eq:GW_Spectrum}, we can notice that the shape of the GW spectrum will depend on the specific choice of the primordial scalar power spectrum. 

\hspace*{1em}\textbf{Red-tilted} First, we consider the nearly scale-invariant primordial curvature power spectrum described by
\begin{equation}
    \Delta_\zeta(k) = A_{\zeta} \left(\frac{k}{k_{*}}\right)^{n_s-1}\,,
\end{equation}
where $A_s = 2.1 \cdot 10^{-9}$ is the amplitude observed at the pivot scale $k_* = 0.05\, {\rm Mpc}^{-1}$ and $n_s = 0.974$ is the tilt of the power spectrum from the combination of Planck, ACT, and DESI~\cite{Planck:2018vyg, ACT:2025fju}.

\hspace*{1em}\textbf{LogNormal}
The curvature power spectrum with a lognormal peak is predicted in hybrid and multi-field models of inflation~\cite{Garcia-Bellido:1996mdl, Garcia-Bellido:2007nns, Kawasaki:2015ppx, Palma:2020ejf}. For example, models with an axion spectator field coupled to the SU(2) gauge fields~\cite{Barnaby:2011qe} lead to the lognormal shape of the power spectrum that can be parametrized by 
\begin{equation}
       \Delta_\zeta(k)  = \frac{A_s}{\sqrt{2\pi\sigma^2}} e^{-\frac{\ln^2{\left(\frac{k}{k_{s}}\right)}}{2\sigma^2}}\,,
\end{equation}
where for the width of the peak at the scale $k_s  =   10^{20} \, {\rm Mpc}^{-1} $ we choose $\sigma=0.1$ and $A_{s}= 10^{-2}$ for the amplitude \cite{Iacconi:2021ltm}.

\hspace*{1em}\textbf{Blue-tilted} The primordial curvature power spectrum is significantly less constrained on smaller scales, where it can deviate from scale-invariance. Therefore, it is interesting to consider also a blue-tilted spectrum of curvature perturbations on small scales~\cite{Gong:2010zf, Ebadi:2023xhq, deKruijf:2024voc}
\[
      \Delta_\zeta(k) =
\begin{cases}
   A_{\zeta} \left(\frac{k}{k_{*}}\right)^{n_s-1},&  k< k_{\rm t.p.} \\
     A_{\zeta}  \left(\frac{k_{\rm t.p.}}{k_{*}}\right)^{n_s-1} \left(\frac{k}{k_{\rm t.p.}}\right)^{n_b-1},              & k\geq k_{\rm t.p.}
\end{cases}
\]
where $n_b = 2$ is the spectral tilt beyond the turning point $k_{\rm t.p.} = 2\cdot 10^{14}{\rm Mpc}^{-1}$, chosen such that the non-linearity scale, defined as $\Delta_{\zeta} = 1$, coincides with $p_{\rm max}$ .

\begin{figure}
    \centering
    \includegraphics[width=.9\linewidth]{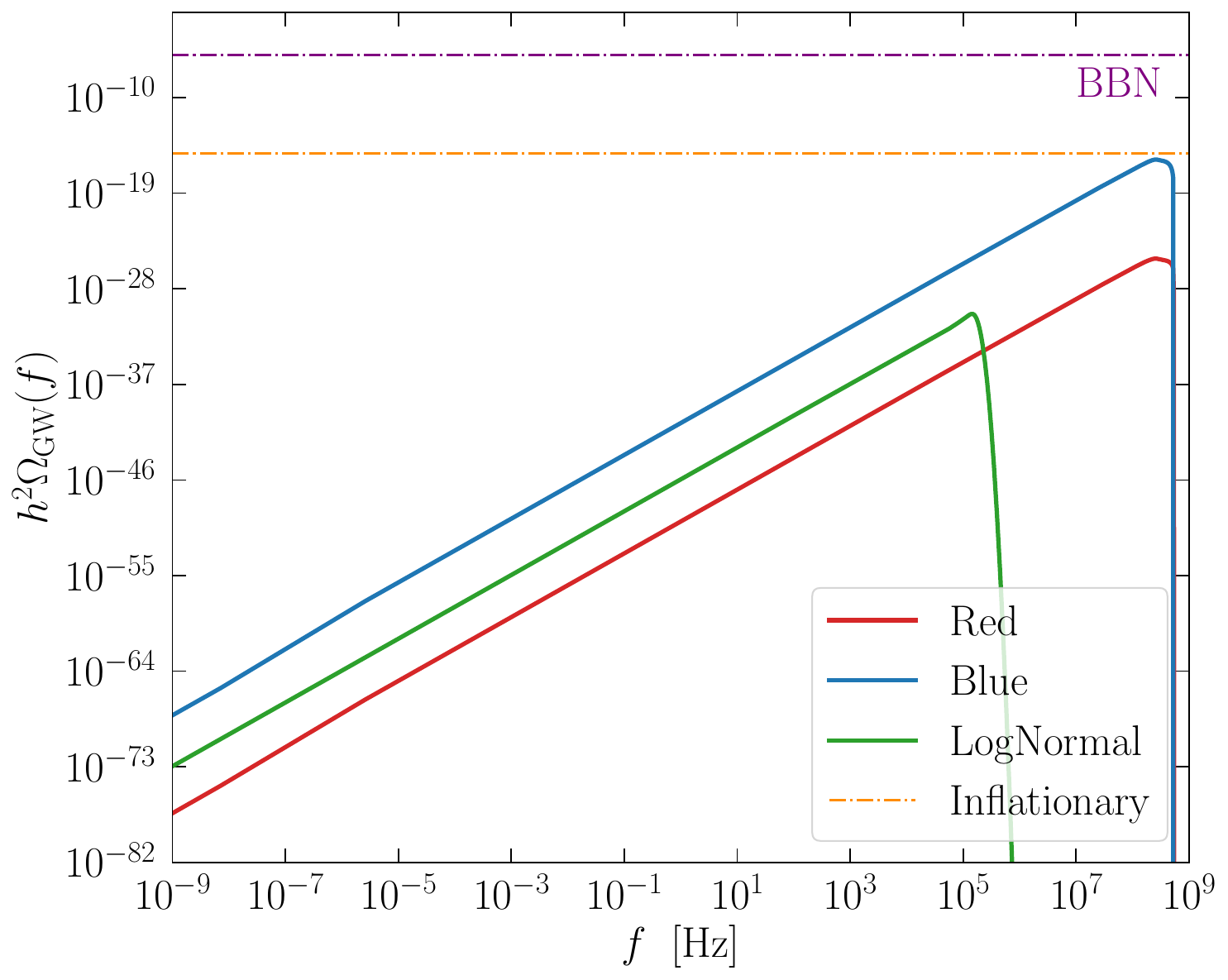}
    \caption{GW spectrum from vacuum fluctuations for different primordial scalar power spectra. We also show in purple the constraint on the total energy density of a GW background at the time of Big Bang Nucleosynthesis (BBN)~\cite{Cielo:2023bqp} and the scale-invariant inflationary GWB~\cite{Caprini:2018mtu} in orange as a reference value. } 
    \label{fig:enter-label}
\end{figure}

We report in Fig. \ref{fig:enter-label}, the resulting GWB spectra today, obtained with the different models. As expected, the magnitude of the GW spectrum strongly depends on the chosen scalar power spectrum. Nevertheless, due to the quartic dependence on the frequency, the maximum value for red- and blue-tilted spectrum is reached at $p_{\rm max}$, while for the lognormal spectrum at $k_s$,
    \begin{equation}
\begin{split}
      h^2\Omega_{\rm GW}^{\rm Red}(p_{\rm max}) &\sim 5.6 \cdot  10^{-26}\,,\\
      h^2 \Omega_{\rm GW}^{\rm Blue}(p_{\rm max}) &\sim 1.2 \cdot 10^{-16}\,,\\
      h^2 \Omega_{\rm GW}^{\rm LN}(k_s) &\sim  4.3 \cdot 10^{-31}\,,
\end{split}
\end{equation}
which is beyond the reach of current and future GW interferometers~\cite{LISACosmologyWorkingGroup:2022jok,Branchesi:2023mws,Abac:2025saz}. However, this makes them a target for the future high-frequency GW detectors~\cite{Aggarwal:2020olq, Aggarwal:2025noe}.  Several innovative proposals are currently being explored, relying on quantum sensing, resonant cavities, opto-mechanical systems, or electromagnetic conversion mechanisms, all of which could in principle probe the GHz frequency band.

Furthermore, other mechanisms such as preheating can increase the GW signal due to the resonant amplification of inflaton fluctuations that gives rise to the enhanced curvature power spectrum~\cite{LISACosmologyWorkingGroup:2025vdz}.
\section{Including the eﬀect of GWs from inflation}
Earlier, we neglected the primordial GWs and assumed a vacuum initial state at the end of inflation. In reality, the initial state at the end of inflation will be a mixed state and the particle creation will be  accompanied by the corresponding stimulated emission~\cite{Agullo:2010ws, Parker:2015nna}. The spectrum of GWs created in the mixed state described by a statistical density $\rho$ is given by 
\begin{equation}
\begin{split}
     \Omega^{\rm tot}_{\rm GW}(k) &=      \Omega^{\rm prim}_{\rm GW}(k)+    \Omega^{\rm BD}_{\rm GW}(k) \left(1+ 2 \Tr[\rho N_{\vec{k}}] \right)\\
     &\simeq \Omega^{\rm prim}_{\rm GW}(k)+    \Omega^{\rm BD}_{\rm GW}(k)\, ,
\end{split} 
\end{equation}
where $\Tr[\rho N_{\vec{k}}]$ is the average number of particles present in mode $\vec{k}$ at the end of inflation. Since in standard models of inflation $\Tr[\rho N_{\vec{k}}] $\footnote{In standard single-field models of inflation $\Tr[\rho N_{\vec{k}}] \propto H_{\rm inf}^2/m_{\rm Pl}^2 .$} is significantly smaller than unity, the total GW spectrum will be the sum of the primordial one and that arising from the Bunch-Davies initial state at $\eta=0$, as considered here.

\section{Conclusions} 

In this paper we have shown that a new cosmological GWB is generated via the CGPP mechanism after inflation during the radiation-dominated epoch, by the scalar perturbations of the metric. We have estimated the power spectrum of these GWs and found that, due to the quartic dependence on frequency, the spectrum reaches its maximum around the GHz frequency range, far above the expected maximum for other astrophysical and cosmological sources of GWs.
Unlike the primordial GWs from inflation that evolve for a long time outside the horizon, our production mechanism occurs recently and on subhorizon scales, therefore one might argue that the quantum signature
of these GWs is more likely to survive to the present day, compared to the primordial signal~\footnote{Still, see~\cite{deKruijf:2024ufs} for an analysis of those scales where the primordial GWs might not have fully decohered, and~\cite{Belfiglio:2025cst} for quantum signatures possibly connected with our mechanism.}.

A natural extension of this work would be to quantify how an early matter-dominated era and a more gradual transition from inflation to the post-inflationary phase modify the GW spectrum. We also plan to explore scenarios in which the pre-existing tensor modes contribute significantly to the signal.

\paragraph{Acknowledgments.}
\noindent
This work is partially supported by the MUR Departments of Excellence grant “Quantum Frontiers” of the Physics and Astronomy Department of Padova University.
NB, SM and GP acknowledge financial support from the COSMOS network (www.cosmosnet.it) through the ASI (Italian Space Agency) Grants 2016-24-H.0, 2016-24-H.1-2018 and 2020-9-HH.0. GP thanks Fondazione Angelo Della Riccia and Fondazione Aldo Gini for financial support.

\vskip 1.5cm

\bibliographystyle{JHEP}
\bibliography{Biblio.bib}

\providecommand{\href}[2]{#2}\begingroup\raggedright\begin{thebibliography}{10}

\bibitem{Parker:1968mv}
L.~Parker, \emph{{Particle creation in expanding universes}},
  \href{https://doi.org/10.1103/PhysRevLett.21.562}{\emph{Phys. Rev. Lett.}
  {\bfseries 21} (1968) 562}.

\bibitem{Parker:1969au}
L.~Parker, \emph{{Quantized fields and particle creation in expanding
  universes. 1.}}, \href{https://doi.org/10.1103/PhysRev.183.1057}{\emph{Phys.
  Rev.} {\bfseries 183} (1969) 1057}.

\bibitem{Parker:1971pt}
L.~Parker, \emph{{Quantized fields and particle creation in expanding
  universes. 2.}}, \href{https://doi.org/10.1103/PhysRevD.3.346}{\emph{Phys.
  Rev. D} {\bfseries 3} (1971) 346}.

\bibitem{Ford:2021syk}
L.H.~Ford, \emph{{Cosmological particle production: a review}},
  \href{https://doi.org/10.1088/1361-6633/ac1b23}{\emph{Rept. Prog. Phys.}
  {\bfseries 84} (2021) } [\href{https://arxiv.org/abs/2112.02444}{{\ttfamily
  2112.02444}}].

\bibitem{Kolb:2023ydq}
E.W.~Kolb and A.J.~Long, \emph{{Cosmological gravitational particle production
  and its implications for cosmological relics}},
  \href{https://doi.org/10.1103/RevModPhys.96.045005}{\emph{Rev. Mod. Phys.}
  {\bfseries 96} (2024) 045005}
  [\href{https://arxiv.org/abs/2312.09042}{{\ttfamily 2312.09042}}].

\bibitem{Capanelli:2024pzd}
C.~Capanelli, L.~Jenks, E.W.~Kolb and E.~McDonough, \emph{{Runaway
  Gravitational Production of Dark Photons}},
  \href{https://doi.org/10.1103/PhysRevLett.133.061602}{\emph{Phys. Rev. Lett.}
  {\bfseries 133} (2024) 061602}
  [\href{https://arxiv.org/abs/2403.15536}{{\ttfamily 2403.15536}}].

\bibitem{Zeldovich:1977vgo}
Y.B.~Zel'dovich and A.A.~Starobinsky, \emph{{Rate of particle production in
  gravitational fields}}, {\emph{JETP Lett.} {\bfseries 26} (1977) 252}.

\bibitem{Birrell:1979pi}
N.D.~Birrell and P.C.W.~Davies, \emph{{Massive Particle Production in
  Anisotropic Space-times}},
  \href{https://doi.org/10.1088/0305-4470/13/6/032}{\emph{J. Phys. A}
  {\bfseries 13} (1980) 2109}.

\bibitem{Cespedes:1989kh}
J.~Cespedes and E.~Verdaguer, \emph{{Particle Production in Inhomogeneous
  Cosmologies}}, \href{https://doi.org/10.1103/PhysRevD.41.1022}{\emph{Phys.
  Rev. D} {\bfseries 41} (1990) 1022}.

\bibitem{Maroto:2000zu}
A.L.~Maroto, \emph{{Primordial magnetic fields from metric perturbations}},
  \href{https://doi.org/10.1103/PhysRevD.64.083006}{\emph{Phys. Rev. D}
  {\bfseries 64} (2001) 083006}
  [\href{https://arxiv.org/abs/hep-ph/0008288}{{\ttfamily hep-ph/0008288}}].

\bibitem{Maleknejad:2024ybn}
A.~Maleknejad and J.~Kopp, \emph{{Gravitational Wave-Induced Freeze-In of
  Fermionic Dark Matter}},  \href{https://arxiv.org/abs/2405.09723}{{\ttfamily
  2405.09723}}.

\bibitem{Maleknejad:2024hoz}
A.~Maleknejad and J.~Kopp, \emph{{Weyl fermion creation by cosmological
  gravitational wave background at 1-loop}},
  \href{https://doi.org/10.1007/JHEP01(2025)023}{\emph{JHEP} {\bfseries 01}
  (2025) 023} [\href{https://arxiv.org/abs/2406.01534}{{\ttfamily
  2406.01534}}].

\bibitem{Garani:2024isu}
R.~Garani, M.~Redi and A.~Tesi, \emph{{Stochastic Dark Matter from Curvature
  Perturbations}},
  \href{https://doi.org/10.1103/PhysRevLett.134.101005}{\emph{Phys. Rev. Lett.}
  {\bfseries 134} (2025) 101005}
  [\href{https://arxiv.org/abs/2408.15987}{{\ttfamily 2408.15987}}].

\bibitem{Garani:2025qnm}
R.~Garani, M.~Redi and A.~Tesi, \emph{{Particle production from
  inhomogeneities: general metric perturbations}},
  \href{https://arxiv.org/abs/2502.12249}{{\ttfamily 2502.12249}}.

\bibitem{Belfiglio:2025chv}
A.~Belfiglio and O.~Luongo, \emph{{Gravitational dark matter production from
  fermionic spectator fields during inflation}},
  \href{https://arxiv.org/abs/2504.04219}{{\ttfamily 2504.04219}}.

\bibitem{NANOGrav:2023gor}
{\scshape NANOGrav} collaboration, \emph{{The NANOGrav 15 yr Data Set: Evidence
  for a Gravitational-wave Background}},
  \href{https://doi.org/10.3847/2041-8213/acdac6}{\emph{Astrophys. J. Lett.}
  {\bfseries 951} (2023) L8}
  [\href{https://arxiv.org/abs/2306.16213}{{\ttfamily 2306.16213}}].

\bibitem{EPTA:2023fyk}
{\scshape EPTA, InPTA:} collaboration, \emph{{The second data release from the
  European Pulsar Timing Array - III. Search for gravitational wave signals}},
  \href{https://doi.org/10.1051/0004-6361/202346844}{\emph{Astron. Astrophys.}
  {\bfseries 678} (2023) A50}
  [\href{https://arxiv.org/abs/2306.16214}{{\ttfamily 2306.16214}}].

\bibitem{Reardon:2023gzh}
D.J.~Reardon et~al., \emph{{Search for an Isotropic Gravitational-wave
  Background with the Parkes Pulsar Timing Array}},
  \href{https://doi.org/10.3847/2041-8213/acdd02}{\emph{Astrophys. J. Lett.}
  {\bfseries 951} (2023) L6}
  [\href{https://arxiv.org/abs/2306.16215}{{\ttfamily 2306.16215}}].

\bibitem{Xu:2023wog}
H.~Xu et~al., \emph{{Searching for the Nano-Hertz Stochastic Gravitational Wave
  Background with the Chinese Pulsar Timing Array Data Release I}},
  \href{https://doi.org/10.1088/1674-4527/acdfa5}{\emph{Res. Astron.
  Astrophys.} {\bfseries 23} (2023) 075024}
  [\href{https://arxiv.org/abs/2306.16216}{{\ttfamily 2306.16216}}].

\bibitem{NANOGrav:2023hvm}
{\scshape NANOGrav} collaboration, \emph{{The NANOGrav 15 yr Data Set: Search
  for Signals from New Physics}},
  \href{https://doi.org/10.3847/2041-8213/acdc91}{\emph{Astrophys. J. Lett.}
  {\bfseries 951} (2023) L11}
  [\href{https://arxiv.org/abs/2306.16219}{{\ttfamily 2306.16219}}].

\bibitem{NANOGrav:2023hfp}
{\scshape NANOGrav} collaboration, \emph{{The NANOGrav 15 yr Data Set:
  Constraints on Supermassive Black Hole Binaries from the Gravitational-wave
  Background}},
  \href{https://doi.org/10.3847/2041-8213/ace18b}{\emph{Astrophys. J. Lett.}
  {\bfseries 952} (2023) L37}
  [\href{https://arxiv.org/abs/2306.16220}{{\ttfamily 2306.16220}}].

\bibitem{EPTA:2023xxk}
{\scshape EPTA, InPTA} collaboration, \emph{{The second data release from the
  European Pulsar Timing Array - IV. Implications for massive black holes, dark
  matter, and the early Universe}},
  \href{https://doi.org/10.1051/0004-6361/202347433}{\emph{Astron. Astrophys.}
  {\bfseries 685} (2024) A94}
  [\href{https://arxiv.org/abs/2306.16227}{{\ttfamily 2306.16227}}].

\bibitem{Franciolini:2023pbf}
G.~Franciolini, A.~Iovino, Junior., V.~Vaskonen and H.~Veermae, \emph{{Recent
  Gravitational Wave Observation by Pulsar Timing Arrays and Primordial Black
  Holes: The Importance of Non-Gaussianities}},
  \href{https://doi.org/10.1103/PhysRevLett.131.201401}{\emph{Phys. Rev. Lett.}
  {\bfseries 131} (2023) 201401}
  [\href{https://arxiv.org/abs/2306.17149}{{\ttfamily 2306.17149}}].

\bibitem{Franciolini:2023wjm}
G.~Franciolini, D.~Racco and F.~Rompineve, \emph{{Footprints of the QCD
  Crossover on Cosmological Gravitational Waves at Pulsar Timing Arrays}},
  \href{https://doi.org/10.1103/PhysRevLett.132.081001}{\emph{Phys. Rev. Lett.}
  {\bfseries 132} (2024) 081001}
  [\href{https://arxiv.org/abs/2306.17136}{{\ttfamily 2306.17136}}].

\bibitem{Ellis:2023tsl}
J.~Ellis, M.~Lewicki, C.~Lin and V.~Vaskonen, \emph{{Cosmic superstrings
  revisited in light of NANOGrav 15-year data}},
  \href{https://doi.org/10.1103/PhysRevD.108.103511}{\emph{Phys. Rev. D}
  {\bfseries 108} (2023) 103511}
  [\href{https://arxiv.org/abs/2306.17147}{{\ttfamily 2306.17147}}].

\bibitem{Vagnozzi:2023lwo}
S.~Vagnozzi, \emph{{Inflationary interpretation of the stochastic gravitational
  wave background signal detected by pulsar timing array experiments}},
  \href{https://doi.org/10.1016/j.jheap.2023.07.001}{\emph{JHEAp} {\bfseries
  39} (2023) 81} [\href{https://arxiv.org/abs/2306.16912}{{\ttfamily
  2306.16912}}].

\bibitem{Figueroa:2023zhu}
D.G.~Figueroa, M.~Pieroni, A.~Ricciardone and P.~Simakachorn,
  \emph{{Cosmological Background Interpretation of Pulsar Timing Array Data}},
  \href{https://doi.org/10.1103/PhysRevLett.132.171002}{\emph{Phys. Rev. Lett.}
  {\bfseries 132} (2024) 171002}
  [\href{https://arxiv.org/abs/2307.02399}{{\ttfamily 2307.02399}}].

\bibitem{Domenech:2024rks}
G.~Dom{\`e}nech, S.~Pi, A.~Wang and J.~Wang, \emph{{Induced gravitational wave
  interpretation of PTA data: a complete study for general equation of state}},
  \href{https://doi.org/10.1088/1475-7516/2024/08/054}{\emph{JCAP} {\bfseries
  08} (2024) 054} [\href{https://arxiv.org/abs/2402.18965}{{\ttfamily
  2402.18965}}].

\bibitem{Bari:2023rcw}
P.~Bari, N.~Bartolo, G.~Dom\`enech and S.~Matarrese, \emph{{Gravitational waves
  induced by scalar-tensor mixing}},
  \href{https://doi.org/10.1103/PhysRevD.109.023509}{\emph{Phys. Rev. D}
  {\bfseries 109} (2024) 023509}
  [\href{https://arxiv.org/abs/2307.05404}{{\ttfamily 2307.05404}}].

\bibitem{Picard:2023sbz}
R.~Picard and K.A.~Malik, \emph{{Induced gravitational waves: the effect of
  first order tensor perturbations}},
  \href{https://doi.org/10.1088/1475-7516/2024/10/010}{\emph{JCAP} {\bfseries
  10} (2024) 010} [\href{https://arxiv.org/abs/2311.14513}{{\ttfamily
  2311.14513}}].

\bibitem{Agullo:2011xv}
I.~Agullo and L.~Parker, \emph{{Stimulated creation of quanta during inflation
  and the observable universe}},
  \href{https://doi.org/10.1007/s10714-011-1220-8}{\emph{Gen. Rel. Grav.}
  {\bfseries 43} (2011) 2541}
  [\href{https://arxiv.org/abs/1106.4240}{{\ttfamily 1106.4240}}].

\bibitem{Ota:2025yeu}
A.~Ota and Y.~Zhu, \emph{{Graviton stimulated emission in squeezed vacuum
  states}},  \href{https://arxiv.org/abs/2504.06539}{{\ttfamily 2504.06539}}.

\bibitem{Ford:1977dj}
L.H.~Ford and L.~Parker, \emph{{Quantized Gravitational Wave Perturbations in
  Robertson-Walker Universes}},
  \href{https://doi.org/10.1103/PhysRevD.16.1601}{\emph{Phys. Rev. D}
  {\bfseries 16} (1977) 1601}.

\bibitem{Caprini:2018mtu}
C.~Caprini and D.G.~Figueroa, \emph{{Cosmological Backgrounds of Gravitational
  Waves}}, \href{https://doi.org/10.1088/1361-6382/aac608}{\emph{Class. Quant.
  Grav.} {\bfseries 35} (2018) 163001}
  [\href{https://arxiv.org/abs/1801.04268}{{\ttfamily 1801.04268}}].

\bibitem{Baumann:2007zm}
D.~Baumann, P.J.~Steinhardt, K.~Takahashi and K.~Ichiki, \emph{{Gravitational
  Wave Spectrum Induced by Primordial Scalar Perturbations}},
  \href{https://doi.org/10.1103/PhysRevD.76.084019}{\emph{Phys. Rev. D}
  {\bfseries 76} (2007) 084019}
  [\href{https://arxiv.org/abs/hep-th/0703290}{{\ttfamily hep-th/0703290}}].

\bibitem{Bartolo:2007vp}
N.~Bartolo, S.~Matarrese, A.~Riotto and A.~Vaihkonen, \emph{{The Maximal Amount
  of Gravitational Waves in the Curvaton Scenario}},
  \href{https://doi.org/10.1103/PhysRevD.76.061302}{\emph{Phys. Rev. D}
  {\bfseries 76} (2007) 061302}
  [\href{https://arxiv.org/abs/0705.4240}{{\ttfamily 0705.4240}}].

\bibitem{Campos:1991ff}
A.~Campos and E.~Verdaguer, \emph{{Production of spin 1/2 particles in
  inhomogeneous cosmologies}},
  \href{https://doi.org/10.1103/PhysRevD.45.4428}{\emph{Phys. Rev. D}
  {\bfseries 45} (1992) 4428}.

\bibitem{Parker:2009uva}
L.E.~Parker and D.~Toms, \emph{{Quantum Field Theory in Curved Spacetime}:
  {Quantized Field and Gravity}}, Cambridge Monographs on Mathematical Physics,
  Cambridge University Press (8, 2009),
  \href{https://doi.org/10.1017/CBO9780511813924}{10.1017/CBO9780511813924}.

\bibitem{Espinosa:2018eve}
J.R.~Espinosa, D.~Racco and A.~Riotto, \emph{{A Cosmological Signature of the
  SM Higgs Instability: Gravitational Waves}},
  \href{https://doi.org/10.1088/1475-7516/2018/09/012}{\emph{JCAP} {\bfseries
  09} (2018) 012} [\href{https://arxiv.org/abs/1804.07732}{{\ttfamily
  1804.07732}}].

\bibitem{Pi:2020otn}
S.~Pi and M.~Sasaki, \emph{{Gravitational Waves Induced by Scalar Perturbations
  with a Lognormal Peak}},
  \href{https://doi.org/10.1088/1475-7516/2020/09/037}{\emph{JCAP} {\bfseries
  09} (2020) 037} [\href{https://arxiv.org/abs/2005.12306}{{\ttfamily
  2005.12306}}].

\bibitem{Perna:2024ehx}
G.~Perna, C.~Testini, A.~Ricciardone and S.~Matarrese, \emph{{Fully
  non-Gaussian Scalar-Induced Gravitational Waves}},
  \href{https://doi.org/10.1088/1475-7516/2024/05/086}{\emph{JCAP} {\bfseries
  05} (2024) 086} [\href{https://arxiv.org/abs/2403.06962}{{\ttfamily
  2403.06962}}].

\bibitem{Kugarajh:2025rbt}
A.A.~Kugarajh, M.~Traforetti, A.~Maselli, S.~Matarrese and A.~Ricciardone,
  \emph{{Scalar-Induced Gravitational Waves in Modified Gravity}},
  \href{https://arxiv.org/abs/2502.20137}{{\ttfamily 2502.20137}}.

\bibitem{Galloni:2022mok}
G.~Galloni, N.~Bartolo, S.~Matarrese, M.~Migliaccio, A.~Ricciardone and
  N.~Vittorio, \emph{{Updated constraints on amplitude and tilt of the tensor
  primordial spectrum}},
  \href{https://doi.org/10.1088/1475-7516/2023/04/062}{\emph{JCAP} {\bfseries
  04} (2023) 062} [\href{https://arxiv.org/abs/2208.00188}{{\ttfamily
  2208.00188}}].

\bibitem{Planck:2018vyg}
{\scshape Planck} collaboration, \emph{{Planck 2018 results. VI. Cosmological
  parameters}},
  \href{https://doi.org/10.1051/0004-6361/201833910}{\emph{Astron. Astrophys.}
  {\bfseries 641} (2020) A6}
  [\href{https://arxiv.org/abs/1807.06209}{{\ttfamily 1807.06209}}].

\bibitem{ACT:2025fju}
{\scshape ACT} collaboration, \emph{{The Atacama Cosmology Telescope: DR6 Power
  Spectra, Likelihoods and $\Lambda$CDM Parameters}},
  \href{https://arxiv.org/abs/2503.14452}{{\ttfamily 2503.14452}}.

\bibitem{Garcia-Bellido:1996mdl}
J.~Garcia-Bellido, A.D.~Linde and D.~Wands, \emph{{Density perturbations and
  black hole formation in hybrid inflation}},
  \href{https://doi.org/10.1103/PhysRevD.54.6040}{\emph{Phys. Rev. D}
  {\bfseries 54} (1996) 6040}
  [\href{https://arxiv.org/abs/astro-ph/9605094}{{\ttfamily
  astro-ph/9605094}}].

\bibitem{Garcia-Bellido:2007nns}
J.~Garcia-Bellido and D.G.~Figueroa, \emph{{A stochastic background of
  gravitational waves from hybrid preheating}},
  \href{https://doi.org/10.1103/PhysRevLett.98.061302}{\emph{Phys. Rev. Lett.}
  {\bfseries 98} (2007) 061302}
  [\href{https://arxiv.org/abs/astro-ph/0701014}{{\ttfamily
  astro-ph/0701014}}].

\bibitem{Kawasaki:2015ppx}
M.~Kawasaki and Y.~Tada, \emph{{Can massive primordial black holes be produced
  in mild waterfall hybrid inflation?}},
  \href{https://doi.org/10.1088/1475-7516/2016/08/041}{\emph{JCAP} {\bfseries
  08} (2016) 041} [\href{https://arxiv.org/abs/1512.03515}{{\ttfamily
  1512.03515}}].

\bibitem{Palma:2020ejf}
G.A.~Palma, S.~Sypsas and C.~Zenteno, \emph{{Seeding primordial black holes in
  multifield inflation}},
  \href{https://doi.org/10.1103/PhysRevLett.125.121301}{\emph{Phys. Rev. Lett.}
  {\bfseries 125} (2020) 121301}
  [\href{https://arxiv.org/abs/2004.06106}{{\ttfamily 2004.06106}}].

\bibitem{Barnaby:2011qe}
N.~Barnaby, E.~Pajer and M.~Peloso, \emph{{Gauge Field Production in Axion
  Inflation: Consequences for Monodromy, non-Gaussianity in the CMB, and
  Gravitational Waves at Interferometers}},
  \href{https://doi.org/10.1103/PhysRevD.85.023525}{\emph{Phys. Rev. D}
  {\bfseries 85} (2012) 023525}
  [\href{https://arxiv.org/abs/1110.3327}{{\ttfamily 1110.3327}}].

\bibitem{Iacconi:2021ltm}
L.~Iacconi, H.~Assadullahi, M.~Fasiello and D.~Wands, \emph{{Revisiting
  small-scale fluctuations in {\ensuremath{\alpha}}-attractor models of
  inflation}}, \href{https://doi.org/10.1088/1475-7516/2022/06/007}{\emph{JCAP}
  {\bfseries 06} (2022) 007}
  [\href{https://arxiv.org/abs/2112.05092}{{\ttfamily 2112.05092}}].

\bibitem{Gong:2010zf}
J.-O.~Gong and M.~Sasaki, \emph{{Waterfall field in hybrid inflation and
  curvature perturbation}},
  \href{https://doi.org/10.1088/1475-7516/2011/03/028}{\emph{JCAP} {\bfseries
  03} (2011) 028} [\href{https://arxiv.org/abs/1010.3405}{{\ttfamily
  1010.3405}}].

\bibitem{Ebadi:2023xhq}
R.~Ebadi, S.~Kumar, A.~McCune, H.~Tai and L.-T.~Wang, \emph{{Gravitational
  waves from stochastic scalar fluctuations}},
  \href{https://doi.org/10.1103/PhysRevD.109.083519}{\emph{Phys. Rev. D}
  {\bfseries 109} (2024) 083519}
  [\href{https://arxiv.org/abs/2307.01248}{{\ttfamily 2307.01248}}].

\bibitem{deKruijf:2024voc}
J.~de~Kruijf, E.~Vanzan, K.K.~Boddy, A.~Raccanelli and N.~Bartolo,
  \emph{{Searching for blue-tilted power spectra in the dark ages}},
  \href{https://doi.org/10.1103/PhysRevD.111.063507}{\emph{Phys. Rev. D}
  {\bfseries 111} (2025) 063507}
  [\href{https://arxiv.org/abs/2408.04991}{{\ttfamily 2408.04991}}].

\bibitem{LISACosmologyWorkingGroup:2022jok}
{\scshape LISA Cosmology Working Group} collaboration, \emph{{Cosmology with
  the Laser Interferometer Space Antenna}},
  \href{https://doi.org/10.1007/s41114-023-00045-2}{\emph{Living Rev. Rel.}
  {\bfseries 26} (2023) 5} [\href{https://arxiv.org/abs/2204.05434}{{\ttfamily
  2204.05434}}].

\bibitem{Branchesi:2023mws}
M.~Branchesi et~al., \emph{{Science with the Einstein Telescope: a comparison
  of different designs}},
  \href{https://doi.org/10.1088/1475-7516/2023/07/068}{\emph{JCAP} {\bfseries
  07} (2023) 068} [\href{https://arxiv.org/abs/2303.15923}{{\ttfamily
  2303.15923}}].

\bibitem{Abac:2025saz}
A.~Abac et~al., \emph{{The Science of the Einstein Telescope}},
  \href{https://arxiv.org/abs/2503.12263}{{\ttfamily 2503.12263}}.

\bibitem{Aggarwal:2020olq}
N.~Aggarwal et~al., \emph{{Challenges and opportunities of gravitational-wave
  searches at MHz to GHz frequencies}},
  \href{https://doi.org/10.1007/s41114-021-00032-5}{\emph{Living Rev. Rel.}
  {\bfseries 24} (2021) 4} [\href{https://arxiv.org/abs/2011.12414}{{\ttfamily
  2011.12414}}].

\bibitem{Aggarwal:2025noe}
N.~Aggarwal et~al., \emph{{Challenges and Opportunities of Gravitational Wave
  Searches above 10 kHz}},  \href{https://arxiv.org/abs/2501.11723}{{\ttfamily
  2501.11723}}.

\bibitem{LISACosmologyWorkingGroup:2025vdz}
{\scshape LISA Cosmology Working Group} collaboration, \emph{{Reconstructing
  primordial curvature perturbations via scalar-induced gravitational waves
  with LISA}}, \href{https://doi.org/10.1088/1475-7516/2025/05/062}{\emph{JCAP}
  {\bfseries 05} (2025) 062}
  [\href{https://arxiv.org/abs/2501.11320}{{\ttfamily 2501.11320}}].

\bibitem{Cielo:2023bqp}
M.~Cielo, M.~Escudero, G.~Mangano and O.~Pisanti, \emph{{Neff in the Standard
  Model at NLO is 3.043}},
  \href{https://doi.org/10.1103/PhysRevD.108.L121301}{\emph{Phys. Rev. D}
  {\bfseries 108} (2023) L121301}
  [\href{https://arxiv.org/abs/2306.05460}{{\ttfamily 2306.05460}}].

\bibitem{Agullo:2010ws}
I.~Agullo and L.~Parker, \emph{{Non-gaussianities and the Stimulated creation
  of quanta in the inflationary universe}},
  \href{https://doi.org/10.1103/PhysRevD.83.063526}{\emph{Phys. Rev. D}
  {\bfseries 83} (2011) 063526}
  [\href{https://arxiv.org/abs/1010.5766}{{\ttfamily 1010.5766}}].

\bibitem{Parker:2015nna}
L.~Parker, \emph{{Creation of quantized particles, gravitons and scalar
  perturbations by the expanding universe}},
  \href{https://doi.org/10.1088/1742-6596/600/1/012001}{\emph{J. Phys. Conf.
  Ser.} {\bfseries 600} (2015) 012001}
  [\href{https://arxiv.org/abs/1503.00359}{{\ttfamily 1503.00359}}].

\bibitem{deKruijf:2024ufs}
J.~de~Kruijf and N.~Bartolo, \emph{{The effect of quantum decoherence on
  inflationary gravitational waves}},
  \href{https://doi.org/10.1088/1475-7516/2024/11/041}{\emph{JCAP} {\bfseries
  11} (2024) 041} [\href{https://arxiv.org/abs/2408.02563}{{\ttfamily
  2408.02563}}].

\bibitem{Belfiglio:2025cst}
A.~Belfiglio, O.~Luongo and S.~Mancini, \emph{{Quantum entanglement in
  cosmology}},  \href{https://arxiv.org/abs/2506.03841}{{\ttfamily
  2506.03841}}.

\end{thebibliography}\endgroup

\end{document}